\documentclass[prl,aps,twocolumn,floats,showpacs]{revtex4}
\usepackage{amsmath}
\usepackage{graphicx}
\usepackage{epsfig}

\newcommand{\ggf}{}
\newcommand{\gf}{}
\newcommand{\be}{\begin{equation}}
\newcommand{\ee}{\end{equation}}
\newcommand{\bea}{\begin{eqnarray}}
\newcommand{\eea}{\end{eqnarray}}

\begin{document}
\title{Exotic Kondo effect from magnetic trimers}

\author{B. Lazarovits$^1$, P. Simon$^2$, G. Zar\'and$^3$, and L. Szunyogh,$^{1,3}$} 
\affiliation{ 
$^1$Center for Computational Materials Science, Vienna University of Technology, A-1060, Gumpendorferstr. 1.a., Vienna, Austria\\
$^2$Laboratoire de Physique et Mod\'elisation des Milieux Condens\'es, CNRS et Universit\'e Joseph Fourier, 38042 Grenoble, France \\
$^3$Theoretical Physics Department, Budapest University of Technology and Economics,Budafoki \'ut 8. H-1521 Hungary 
}
\date{\today}

\begin{abstract} 
Motivated by the recent experiments of Jamneala {\em et al.} [T. Jamnaela {\em et al.}, 
Phys. Rev. Lett. {\bf 87}, 256804 (2001)], by combining
{\em ab-initio} and renormalization group methods, we study 
the strongly correlated state of a Cr trimer deposited on 
gold.  Internal orbital fluctuations of the trimer lead to huge increase 
of $T_K$ compared to the single ion Kondo temperature explaining the experimental 
observation of a zero-bias anomaly for the trimers. The strongly correlated 
state seems to belong to a new, yet hardly explored class of 
non--Fermi liquid fixed points. 
\end{abstract}
\pacs{75.20.Hr, 71.27.+a, 72.15.Qm}

\maketitle

In recent years, atomic scale resolution 
Scanning Tunneling Microscopy (STM) proved to be a spectacular tool 
to probe the local density of states
around Kondo impurities adsorbed on a metallic surface \cite{schneider98,madhavan98}. 
Experiments were performed with single
Ce atoms on Ag \cite{schneider98}, and Co atoms on Au \cite{madhavan98} and Cu surfaces \cite{eigler00,knorr02}. 
When the STM tip is placed directly on the top of the magnetic adatom, a sharp resonance 
appears in the differential conductance at low bias and disappears when the STM tip is moved away from the
 impurity or when the substrate temperature is raised above the Kondo temperature $T_K$. 
This zero bias anomaly appears as the main signature of the  Kondo effect and results from the screening 
of the adatom spin by the surrounding (bulk and surface) conduction electrons.
The precise line shape of the resonance can well be understood in terms of a 
Fano resonance, an interference phenomenon occurring because of two possible tunneling channels: a direct channel between the tip and the 
impurity and a second channel between the tip and the surface \cite{madhavan98,schiller00,ujsaghy00,plihal01}.

\begin{figure}[tb]
\begin{tabular}{ccc}
\epsfxsize3.0cm
\epsfbox{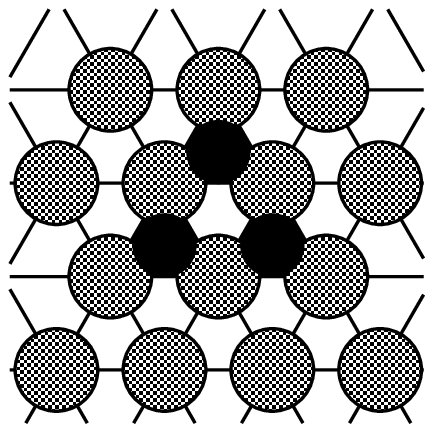}
&& 
\epsfxsize3.2cm
\epsfbox{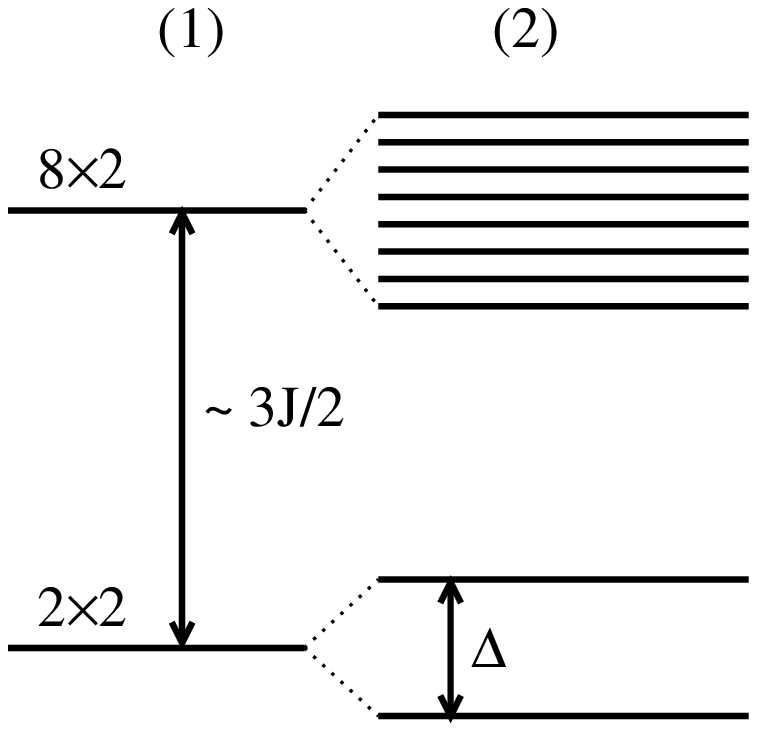}
\\
Fig.~a&&
Fig.~b
\end{tabular}
\vskip0.1cm
\caption{\label{fig:layout+levels}
Fig.~a: Top view of the equilateral Cr cluster on the Au(111) surface. 
Au and Cr atoms are indicated by grey and black circles, respectively. 
Fig.~b: Eigenstate structure of $H_{\rm spin}$  of the Cr cluster (1) in the absence and (2) in the 
presence of spin-orbit coupling. }
\end{figure}

\vspace{-2pt}
The manipulation of single atoms on top of a surface with an STM tip has also been
proven useful to build clusters of atoms with well-controlled interatomic distances. 
For example, Manoharan {\em et al.} \cite{eigler00} manufactured an elliptical 
quantum corral of Co adatoms, and found that,
when an extra Co adatom is placed at 
one focus of the elliptical corral, a ``mirage'' of the 
Kondo resonance can also be observed at the other focus. 
Another 
intriguing result was found recently 
for Cr trimers on a gold surface by Jamneala  {\em et al.} 
\cite{crommie01}. 
Whereas isolated Cr monomers
or dimers display featureless signals in STM spectra at $T=7K$, Cr trimers 
exhibit two distinct electronic states depending on the atomic
positions: 
a sharp Kondo resonance of width $T_K\sim 50K$ was found for an
equilateral triangle, while the STM signal of an isosceles triangle 
did not show any particular feature.
Furthermore, the Cr trimers were reversibly switched from one state to another. 
As schematically depicted in Fig.~\ref{fig:layout+levels}, 
Cr atoms forming an equilateral triangle are expected to occupy 
nearest neighbor sites on the gold 
surface (Au(111)), 
allowing therefore geometric frustration \cite{crommie01}. 
Such a compact magnetic nanocluster is of very much theoretical interest
due to the interplay between Kondo physics and magnetic frustration that 
may generate internal orbital fluctuations. 
This system can indeed be regarded as the smallest and simplest 
frustrated Kondo lattice and, as we will see, embodies a very rich behavior.
We show in this Letter that internal orbital fluctuations in the Cr equilateral trimer lead to a 
huge increase of the Kondo temperature in agreement with the experiment and 
{\gf we argue that} the low energy physics of the system is 
governed by a new  non--Fermi liquid fixed point \cite{ingersent}.

In this Letter we shall present a careful study of 
the strongly correlated state of the Cr trimer depicted in 
Fig.~\ref{fig:layout+levels}.  In order to construct an effective Hamiltonian for this system, 
we first performed {\em ab-initio} calculations \cite{LSW02} to study the 
electronic structure 
of the Cr ions forming the cluster, and verified that the Cr ions are within a very 
good approximation in $d^5$ spin $S_{\rm Cr} = \tilde S=5/2$ states, and display 
relatively small valence fluctuations. Under these conditions, the Hamiltonian describing 
the nanocluster can well be approximated as 
\begin{equation}
H =  H_{\rm spin} + \frac G2 \sum_{i,\sigma,\sigma'} 
{\vec S_i} \psi^\dagger_{i\sigma} {\vec \sigma}_{\sigma\sigma'}  \psi_{i\sigma'}\;,
\label{eq:H}
\end{equation}
where $H_{\rm spin}$ describes the interaction between the Cr spins 
$ {\vec S}_i$ at sites $i=1,..,3$, 
and $G$ denotes the Kondo coupling between each Cr  spin and the conduction electrons 
in the  substrate. Note that Eq.~(\ref{eq:H}) incorporates only  the exchange generated by the most strongly 
hybridizing $d$-state. This approximation is justified by the observation that the Kondo effect is exponentially 
sensitive to the strength of hybridization, and therefore hybridization with other $d$-states is not expected to 
change our results.  Correspondingly, it is sufficient to consider only the hybridization with a 
single conduction electron state, $\psi_i$, leading to Eq.~(\ref{eq:H}).
Note, however, that these states may overlap with each-other and in general 
$\{ \psi_i, \psi^\dagger_j\}\ne \delta_{i,j}$. As we  demonstrate later by {\em ab-initio} calculations, 
the effective spin-spin interaction $H_{\rm spin}$ in Eq. (\ref{eq:H})  
is rather  well approximated by a  dipolar term 
$H_{\rm spin}\approx H_{\rm dipol} = J \sum_{(i,j)} {\vec S}_i {\vec S}_j$, where the coupling $J\approx 1600 {\rm K}$ 
turns out to be antiferromagnetic.

Since the Kondo temperature $T_K\sim 50{\rm K}$ generated by $G$ 
is expected to be much smaller than the exchange coupling $J$,
and we  only wish to describe the physical behavior of the cluster around the low energy scale $\sim T_K$,
we shall first diagonalize $H_{\rm spin}$ and construct an effective Hamiltonian
 to describe quantum fluctuations of the cluster spins. 
The The low energy section of the spectrum of $H_{\rm dipol} $ is sketched in 
Fig.~\ref{fig:layout+levels}b:
There are {\em two} different ways to construct 
states with total spin $S=1/2$ that minimize $H_{\rm dipol}$. 
As a consequence, 
the ground state 
is fourfold degenerate, the extra degeneracy being associated with a two-dimensional representation 
of the  $C_{3v}$ symmetry of the cluster. These four spin states can thus be labeled as 
$|\sigma, \mu\rangle$, where $\sigma = \uparrow,\downarrow$ denotes the spin and $\mu=\pm$ is a chiral index:
$C_3|\sigma, \mu\rangle = e^{i\mu 2\pi/3}|\sigma, \mu\rangle$. Note that the four states remain degenerate 
as long as $H_{\rm spin}$ is SU(2)-invariant, and only spin-orbit coupling related effects discussed later 
can split their degeneracy.

 Next , we shall obtain the effective Hamiltonian within the subspace $\{|\sigma, \mu\rangle\}$. 
To do this we first  construct the states $|\sigma, \mu\rangle$ explicitly using Clebsch-Gordan 
coefficients and then  evaluate the matrix elements of the Kondo exchange, 
$H_K \equiv \frac G2 \sum_{i,\sigma,\sigma'} 
{\vec S_i} \psi^\dagger_{i\sigma} {\vec \sigma}_{\sigma\sigma'}  \psi_{i\sigma'}$ within this subspace in order to obtain an effective Hamiltonian $H_{\rm eff}$.
The effect of virtual transitions into the high energy subspace on the 
effective low energy couplings is neglected assuming $G/J\ll 1$ 
To express the effective Hamiltonian $H_{\rm eff}$ in a simple form we
define the following Fermionic fields, $\psi_{\sigma\mu} \equiv \frac {1}{\sqrt{3}} \sum_j
e^{i\;\mu j\; 2\pi/3} \psi_{j\sigma}$, where now $\mu$ takes three possible values, $\mu = 0, \pm$. 
After tedious and lengthy algebraic manipulations, the effective Hamiltonian takes a rather simple form 
 in this notation,
\begin{eqnarray} 
H_{\rm eff} & = & {\frac{G}{6}} \Bigl[ {\vec S}\; \psi^\dagger {\vec \sigma} \psi \nonumber \\
& - & (2\tilde S+1) \bigl({\vec S}\; T^+ \;\psi^\dagger {\vec \sigma} \tau^- \psi + {\rm h.c.} \bigr)\Bigr]\;,
\label{eq:H_eff}
\end{eqnarray}
where ${\vec S}$ denotes a spin 1/2 operator acting on the spin indices of $|\sigma,\mu\rangle$, and the 
orbital pseudospin operators $T^\pm$ are standard operators raising/lowering the chiral spin $\mu$.  
The operators $\tau^\pm$ in Eq.~(\ref{eq:H_eff}) change 
the angular quasi momentum of the conduction electrons by one unit,
$\tau^\pm_{\mu,\mu'} = {\delta}^{(3)}_{\mu,\mu'\pm 1}$, where $\delta^{(3)}$ 
denotes the Kronecker delta function modulo 3. Note that $H_{eff}$ depends on the magnitude of the initial local spins enhancing the
second term by a factor $2\tilde S+1$.  

To determine the low-energy dynamics of the system and the Kondo scale $T_K$ associated with 
the binding energy of its strongly correlated  state, we carried out a perturbative 
renormalization group (RG) analysis of the system.   
Within the RG we integrate out conduction electrons with large energy
and take their effect into account by changing the coupling constants of the original model. 
Thereby, the original bandwidth $D_0$ is gradually reduced to smaller and smaller values $D$. 
Two stages must be distinguished in course of the RG procedure: (a) For $D\gg J$ the neighboring Cr spins 
behave as {\em independent} spins. (b) For  $D\ll J$ on the other 
hand the Cr spins are tied  together by their exchange interaction, 
and the effective Hamiltonian~(\ref{eq:H_eff}) can be used. 
In the first regime only $G$ is renormalized  according to the 
usual scaling equation\cite{hewson}:
\begin{equation}
\frac{d G}{d l} = \varrho_0 G^2 -  \frac12 \varrho_0^2 G^3 \;\phantom{nnnn}(D \gg J),
\end{equation}  
where $l = {\rm ln}(D_0/D)$ is the scaling parameter and $\varrho_0$ denotes the local density of 
states, and is related to the propagator of the field $\psi_{\sigma j}$ through 
$\langle \psi_{\sigma j}(t) \psi_{\sigma j}^\dagger(0)\rangle \approx -i \varrho_0/t$.

At energies (time scales) below $J$, we can thus  use the effective 
Hamiltonian~(\ref{eq:H_eff}), 
but with a conduction electron cut-off reduced to $D = \tilde D_0 \sim 3J/2$ and the coupling 
$G$ replaced by  an effective coupling $G\to \tilde G \equiv G(l = {\rm ln}(D_0/\tilde D_0))$.  
In this regime, however, the RG analysis becomes more complicated,
and additional terms are generated in the Hamiltonian. 
 To allow for the generation of these terms  we first introduce a very general Hamiltonian of the 
form: 
\begin{equation}
H \equiv  \sum_{\alpha,\beta = 1,..,4} \sum_{p,q=1,..6} |\alpha\rangle\langle \beta | \;
\psi_p^\dagger  V^{\alpha \beta}_{p q} \psi_q\;,
\end{equation}
where $\alpha,\beta = \{\sigma=\pm,\mu=\pm\}$ and $p,q = \{\sigma=\pm,\mu=\pm,0\}$ denote composite indices 
referring to the  ground state multiplet  and the conduction electrons, respectively. 
To obtain $T_K$ we solved numerically the RG equations derived in 
Ref.~\cite{zarand96} for this general model. 
The initial values of the couplings $ V^{\alpha \beta}_{p q}$ can easily
be determined from 
Eq.~(\ref{eq:H_eff}) at $D =  \tilde D_0$. However, special care is needed to 
define the dimensionless couplings entering the scaling equations: 
since $\{ \psi_i, \psi^\dagger_j\}\ne \delta_{i,j}$,
the off-diagonal-correlation function decays in general as 
$\langle \psi_{\sigma 1}(t) \psi_{\sigma 2}^\dagger(0)\rangle \approx -i \alpha \varrho_0/t$, 
with an overlap parameter $\alpha < 1$. 
As a result, the density of states in the electronic 
channels depends on the chiral index: $\varrho_\mu = \varrho_0 (1 + 2 \alpha)$ for $\mu = 0$ and 
 $\varrho_\mu = \varrho_0 (1 - \alpha)$ for $\mu = \pm $. These density of states 
enter the dimensionless couplings of Ref.~\cite{zarand96} 
 as $v^{\alpha\beta}_{pq} \equiv \sqrt{\varrho_p \varrho_q} \;
V^{\alpha\beta}_{pq}$. {\ggf 
The parameter $\alpha$ can be estimated using a free electron 
model \cite{Libero},  but a simple tight binding model can also be used to 
estimate it \cite{Bence_next}, and is typically in the range $0.2 < \alpha < 0.6$.}

The advantage of the RG method discussed above is that it sums up 
systematically leading and subleading logarithmic corrections, and 
provides an unbiased tool to obtain $T_K$, even without knowing the structure of 
the fixed point Hamiltonian. This is in sharp contrast with a variational study 
as the one developed by Kudasov and Uzdin for this problem
\cite{kudasov02}, which only gives an upper 
bound on $T_K$, gets even the exponent of Kondo temperature incorrectly 
(see {\em e.g.} in Ref.~\cite{hewson}), and 
the result of which depends essentially on the specific variational ansatz made. 
Furthermore, the perturbative renormalization group 
enables one to gain some insight into the  structure and symmetry of the fixed point 
Hamiltonian describing the  physics of Eq.~(\ref{eq:H}) below  
the Kondo scale.

\begin{figure}[tb]
\begin{center}
\epsfxsize7cm
\epsfbox{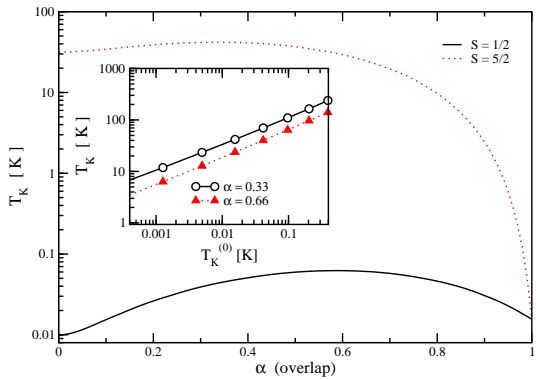}
\end{center}
\vskip0.1cm\caption{\label{fig:TK's}
{\ggf Kondo temperature as a function of the overlap parameter $\alpha$
for a cluster 
formed of atoms with spin  $\tilde S = 5/2$ and spin $\tilde S = 1/2$.
We used $\tilde D_0 = 5000 {\rm \;K}
\approx 3J$ and a dimensionless coupling $G=0.09$. 
The inset shows in a log-log plot the Kondo temperature of the trimer as a function 
of the single Cr's Kondo temperature for two overlap paraemetrs $\alpha=0.33$ and $\alpha=0.6$.}
}
\end{figure}

The Kondo temperature can be 
defined as the energy scale at which the {\ggf 
norm of the effective couplings, $N \equiv (\sum_{\alpha,\beta, p,q} |v^{\alpha\beta}_{pq}|^2)^{1/2}$
reaches the strong coupling limit ($N(D=T_K)\equiv 0.7$). }
For $\alpha=1$ only the field $\psi_{\mu=0}$ couples to the cluster, and the scaling 
equations reduce to those of an isolated  Cr spin. Therefore, in this limit 
$T_K$ is the same as that of an isolated Cr ion, $T_K^{(0)}$. 
{\ggf In the  inset of Fig.~\ref{fig:TK's} we show the Kondo temperature of the 
trimer as a function of the Kondo temperature of a single Cr ion for two typical values 
of the overlap parameter $\alpha$ and for an intermediate cut-off 
$\tilde D_0 = 5000K \sim 3J$. The trimer's Kondo temperature is orders of magnitude larger 
than that of the single Cr ion, and for  $ 0.001 K < T_K^{(0)}< 0.1 K$ (which is in agreement 
with the small bulk Kondo temperature of Cr in Au \cite{T_KCr}), yields a trimer  Kondo scale
$10 K < T_K < 100  K$.  We also find that 
$T_K \approx C(\alpha)\; \sqrt{\tilde D_0\; T_K^{(0)}}$, with $C(\alpha)$ a constant of the order 
of unity.} This increase is clearly due to the presence of orbital fluctuations, 
and gives a natural explanation to the fact that the Kondo effect has not been observed
experimentally for a single Cr ion on Au.\cite{crommie01} Note, however, that 
this increase is peculiar to the 
case of large cluster spins $\tilde S$, and no such a dramatic increase appears for spin $\tilde S = 1/2$ 
trimers, in contrast to the results of Ref.~\cite{kudasov02}. 
{\ggf This important difference of 
behavior between 
$\tilde S=1/2$ and $\tilde S=5/2$ is highlighted in  
Fig.~\ref{fig:TK's},  where on a log scale we displayed
 $T_K$ as a function of the overlap parameter for the  specific choice, $G = 0.09$, 
corresponding to a trimer Kondo scale  $\sim 40 {\rm\; K}$ and 
$T_K^{(0)}\approx 0.01 {\rm \; K}$.} 

Distorting the equilateral trimer configuration, 
we clearly lift the 4-fold ground state degeneracy of $H_{\rm spin}$ and therefore 
suppress this 
Kondo cluster effect and no enhancement of the Kondo temperature is expected. 
Geometric distortion in our formalism appears in fact as a strong orbital magnetic field. 
This gives a natural explanation why the isosceles  
Cr trimer displayed no Kondo resonance at the experimental temperature $T=7K$.

The RG flows also allow us {\ggf to gain information on}  the {\gf symmetry}  of the 
low energy fixed point, governing  the $T\to 0$ dynamics of the cluster, 
{\ggf since dynamically generated  symmetries typically show up already 
within the perturbative RG.}  
Surprisingly enough, {\gf analyzing the structure and algebraic properties 
of the fixed point couplings $v^{\alpha\beta}_{pq}$},
we find that it is {\em neither} the familiar $SU(4)$ Fermi 
liquid fixed point \cite{su4} {\em nor} the two-channel Kondo fixed point \cite{Millis}. 
{\gf We find instead that the fixed point Hamiltonian takes the following 
from:
\bea
H_{\rm fp} &=& \alpha \;{\vec S} \vec \sigma 
+ \beta \;{\vec S} \vec \sigma \; L_z^2   + \gamma \;T^z L^z
\nonumber
\\
&+&  \delta \;{\vec S} \vec \sigma \; (T^- L^+ + T^+ L^-) \;,
\eea
where for simplicity we suppressed the fermion fields. The operators $L^\pm$ and 
$L^z$ above denote standard $L=1$ angular momentum matrices acting on the 
fermionic orbital spin. While this Hamiltonian clearly has an enlarged 
U(1) symmetry [replacing the original $C_{3}$ symmetry], it remains nevertheless 
anisotropic in orbital space. This structure allows us to identify this fixed point 
with the non-Fermi liquid fixed point  identified first 
in Ref.~\cite{ingersent} using the numerical renormalization group method for the simpler problem 
of a spin 1/2 model trimer.
A supplementary strong coupling analysis can be used to support that this 
fixed point is indeed at finite couplings and must therefore be of non-Fermi 
liquid character. We also find that this fixed point is  unstable to general hermitian 
perturbations (another indication of unstable fixed points),  
under which it flows to the familiar stable SU(4) Fermi liquid fixed point\cite{su4}. 
On the other hand,} we could not find  additional terms respecting  
$SU(2)\times C_{3v}$  symmetry that would render the strange fixed point unstable. 
The analysis of the  physical properties of this non-Fermi liquid fixed point 
is beyond the scope of the present paper,  but the  thermodynamical, transport, and 
tunneling (STM) properties are expected to show anomalous scaling properties.
\cite{ingersent,ludwig} 

\begin{figure}[tb]
\begin{center}
\epsfxsize7cm
\epsfbox{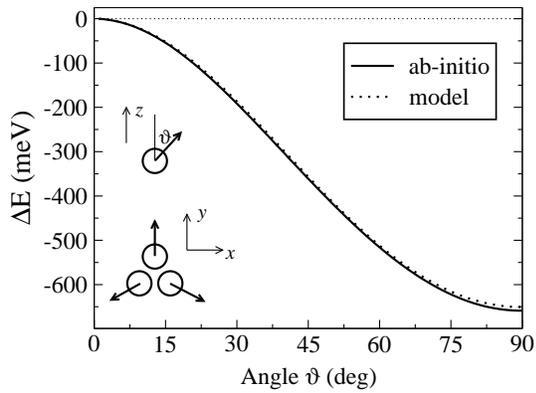}
\end{center}
\vskip0.1cm\caption{\label{fig:comp}
Energy of the Cr trimer as a function of the direction of magnetization
computed using a relativistic  {\em ab-initio} method (full line) and
the effective model, Eq.~(\ref{eq:EOmega}),  
(dotted line) for the particular configurations
sketched in the inset.  In this example only the angle $\vartheta$ with 
respect to the $z$ axis is varied (upper inset).  The lower inset shows the 
projections of the magnetization onto the surface ($xy$ plane). 
}
\end{figure}

So far we fully neglected the effect of  spin-orbit coupling. However, 
spin-orbit coupling removes the four-fold degeneracy of the ground state of 
$H_{\rm spin}$, and splits it up 
into two Kramers doublets. To build a consistent picture (and also to have a 
non-Fermi liquid regime)  this splitting $\Delta$ must be smaller than $T_K$. 
In order to estimate the splitting 
we performed relativistic {\em ab-initio} calculations \cite{LSW02} 
by fixing the orientation of 
 Cr spins, ${\vec S}_j \to S {\vec \Omega}_j$, and determined 
 the energy of the cluster
as a function of the spin orientations, $E[ \{ {\vec \Omega}_j\}]$. 
Details of this calculations will be published elsewhere.\cite{Bence_next}
The energy of the cluster 
can excellently be approximated by the following expression,
\begin{equation} \label{eq:EOmega}
E[\{ {\vec \Omega}_j\}] = S^2 \sum_{i,j=1}^3 \sum_{\alpha,\beta=x,y,z}
J_{ij}^{\alpha\beta} \Omega_i^\alpha \Omega_j^\beta
+ H_{ Q} \quad . 
\end{equation}
The dominant term of the interaction turns out to be 
the simple SU(2) invariant exchange interaction studied in the first part of the paper.
The term $H_{Q}$ above contains SU(2) invariant quadrupolar couplings and 
three-spin 
interactions. While these turn out to be of the same order of magnitude as the anisotropy of the 
exchange interaction, $H_Q$ does not lead to a splitting of the ground state degeneracy
in contrast to anisotropy terms. 
In Fig.~\ref{fig:comp} the {\em ab-initio} values of the energy 
of the cluster are compared with 
energies of the effective Hamiltonian in Eq.~(\ref{eq:EOmega})  
for a particular configuration of the Cr spins.
Having determined the exchange couplings above, we performed first order perturbation 
theory within the degenerate subspace of the trimer to obtain $\Delta \approx 20 {\rm K}$, 
which is indeed less than the experimentally observed Kondo temperature, $T_K \sim 50 {\rm K}$. 
This splitting will ultimately destroy the non-Fermi liquid properties predicted above, 
however, the energy scale $\Delta^*$ at which this happens is expected to be smaller than 
$\Delta$ similar to the two-channel Kondo model, where $\Delta^* \sim \Delta^2/T_K$.


{\gf For Cr on gold the ratio $T_K/\Delta^*$ is  probably too small to observe the 
non-Fermi liquid physics. However,     
it should be possible to observe 
it in other systems: }
Cr on Ag,  {\em e.g.}, is a promising candidate since the lattice constant 
 of Ag is about the same as that of Au (important to have in 
antiferromagnetic coupling between  
magnetic ions),  while the spin-orbit coupling is much weaker. Our calculations show 
that in this case $\Delta\sim 1{\rm K}$, and a much wider non-Fermi liquid range may be 
accessible.  
In this case the atoms would be replaced by 
quantum dots. The great advantage of such a device would be that (1) it would be highly tunable
and (2) it would allow for {\em transport} measurements, though it is not easy to guarantee the perfect 
symmetry of the device.

This research has been supported
by NSF-MTA-OTKA Grant No. INT-0130446, Hungarian Grants No. OTKA
T038162, T046267, and T046303, and the European 'Spintronics' RTN
HPRN-CT-2002-00302.  G.Z. has been supported by the Bolyai Foundation.
B.L. and L.S. were also supported by the Center for Computational Materials
Science (Contract No. GZ 45.531), and the Research and Technological 
Cooperation Project between Austria and Hungary (Contract No. A-3/03).


\vspace{-0.3cm}
\begin{thebibliography}{30}
 \vspace{-.3cm}
\bibitem{schneider98} J. Li, W.-D. Schneider, R. Berndt, and B. Delley, Phys. Rev. Lett. {\bf 80}, 2893 (1998).
\bibitem{madhavan98} V. Madhavan, W. Chen, T. Jamneala, M. F. Crommie, and N. S. Wingreen, Science {\bf 280}, 567 (1998); Phys. Rev. {\bf B 64}, 165412 (2001).
\bibitem{eigler00} H. C. Manoharan, C. P. Lutz, D. M. Eigler, Nature {\bf 403}, 512 (2000).
\bibitem{knorr02} N. Knorr, M. A. Schneider, L. Diekhoener, P. Wahl, and K. Kern, Phys. Rev. Lett. {\bf 88}, 096804 (2002).
\bibitem{schiller00} A. Schiller and S. Hershfield, Phys. Rev. {\bf B 61}, 9036 (2000).
\bibitem{ujsaghy00} O. \'Ujs\'aghy, J. Kroha, L. Szunyogh, and A. Zawadowski, Phys. Rev. Lett. {\bf 85}, 2557 (2000).
\bibitem{plihal01} M. Plihal and J. W. Gadzuk, Phys. Rev. {\bf B 63}, 085404 (2001).
\bibitem{crommie01} T. Jamneala, V. Madhavan, and M. F. Crommie, Phys. Rev. Lett. {\bf 87}, 256804 (2001).
\bibitem{ingersent} B. C. Paul and K. Ingersent, cond-mat/9607190 (1996), unpublished.

\bibitem{Libero} {\ggf V. L. Libero and L. N. Oliveira, Phys. Rev. Lett. {\bf 65}, 2042 (1990).}
\bibitem{LSW02} B. Lazarovits, L. Szunyogh, and P. Weinberger, Phys. Rev. B {\bf 65}, 104441 (2002).
\bibitem{hewson} A.C.~Hewson, {\it The Kondo Problem to Heavy Fermions}
  (Cambridge University Press, Cambridge, UK, 1993).
\bibitem{zarand96} G. Zar\'and, Phys. Rev. Lett. {\bf 77}, 3609 (1996).
\bibitem{kudasov02} Yu. B. Kudasov and V. M. Uzdin, Phys. Rev. Lett. {\bf 89}, 276802 (2002).
\bibitem{su4} L. Borda, G. Zar\'and, W. Hofstetter, B. I. Halperin, and J. von Delft, Phys. Rev. Lett. {\bf 90}, 026602 (2003); G. Zar\'and, A. Brataas and D. Goldhaber-Gordron, Solid State Com. {\bf 126}, 463 (2003); K. Le Hur and P. Simon, Phys. Rev. {\bf B 67}, 201308R (2003). 
\bibitem{T_KCr} {\ggf C. Rizzuto, Rep. Prog. Phys. {\bf 37}, 147 (1974).}
\bibitem{Millis} N. Shah and A. J. Millis,  Phys. Rev. Lett. {\bf 91}, 147204 (2003).
\bibitem{ludwig} K. Ingersent, A.W.W. Ludwig, and I. Affleck, cond-mat/0505303.
\bibitem{Bence_next} {\ggf B. Lazarovits {\em et al.}, under preparation.}

\end{thebibliography}
\end{document}